\documentclass[twocolumn,preprintnumbers,amssymb,amsmath,superscriptaddress,nofootinbib,groupedaddress]{revtex4}
\pdfoutput=1
\usepackage[dvips]{graphicx}
\usepackage{dcolumn} 
\usepackage{bm}
\usepackage{epsfig,amsmath,amssymb,verbatim,mathrsfs,array,layout,textcomp,amssymb,latexsym,slashed}
\usepackage{color}
\usepackage{hyperref}

\input epsf.tex
\newcommand{\beq}{\begin{eqnarray}}
\newcommand{\eeq}{\end{eqnarray}}

\def\ltap{\ \raise.3ex\hbox{$<$\kern-.75em\lower1ex\hbox{$\sim$}}\ }
\def\gtap{\ \raise.3ex\hbox{$>$\kern-.75em\lower1ex\hbox{$\sim$}}\ }

\def\tr{{\rm\ Tr}}

\def\eg{{\it e.g.}}
\def\ie{{\it i.e.}}

\def\be{\begin{equation}}
\def\ee{\end{equation}}
\def\bea{\begin{eqnarray}}
\def\eea{\end{eqnarray}}

\pdfoutput=1

\newcommand{\hc}{{\rm h.c.}}

\def\ma5{{\tt MadAnalysis5\;}}

\def\ma5{{\tt MadAnalysis5\;}}

\begin{document} 


\title{
Bounding Wide Composite Vector Resonances at the LHC
}
\author{Daniele Barducci}
\email{barducci@lapth.cnrs.fr}
\author{C\'edric Delaunay}
\email{delaunay@lapth.cnrs.fr}
\affiliation{LAPTh, Universit\'e Savoie Mont Blanc, CNRS B.P. 110, F-74941 Annecy-le-Vieux, France}
\preprint{\scriptsize LAPTH-061/15\vspace*{.1cm}\\}
\vskip .05in

\begin{abstract}
\vskip .05in
In composite Higgs models (CHMs), electroweak precision data generically push colourless composite vector resonances to a regime where they dominantly decay  into pairs of light top partners. This greatly attenuates their traces in canonical collider searches, tailored for narrow resonances promptly decaying into Standard Model final states. By reinterpreting the CMS same-sign dilepton (SS2$\ell$) analysis at the Large Hadron Collider (LHC), originally designed to search for top partners with electric charge $5/3$, we demonstrate its significant coverage over this kinematical regime. We also show the reach of the 13 TeV run of the LHC, with various integrated luminosity options, for a possible upgrade of the SS2$\ell$ search. The top sector of CHMs is found to be more fine-tuned in the presence of colourless composite resonances in the few TeV range.
\end{abstract}

\maketitle


\section{Introduction}\label{intro}

Composite Higgs models (CHMs), where the Higgs state is realised as a pseudo Nambu--Goldstone boson (pNGB), remain an attractive dynamical explanation of a stable Higgs mass parameter around the weak scale~\cite{Kaplan:1983fs}.
Within this framework, the Higgs field is assumed to be part of a new strong sector beyond the Standard Model (SM) with a cut-off scale in the $5-10\,$TeV range. This solves the so-called big hierarchy problem of the SM, while the remaining little hierarchy between the Higgs mass and the cut-off scale is naturally ensured by the pNGB nature of the Higgs field. More practically, the quadratically divergent contributions to the Higgs mass parameter from SM states are cancelled by contributions arising from the new strong sector resonances.
The naturalness of the weak scale, together with a relatively light Higgs boson, then  requires fermionic resonances coupled to the SM top quark to have a mass below $1\,$TeV~\cite{Matsedonskyi:2012ym,Redi:2012ha,Pomarol:2012qf}. These so-called top partners are actively searched at the Large Hadron Collider (LHC), where current bounds have reached the $700-900\,$GeV mass range~\cite{Aad:2015kqa,Khachatryan:2015oba}, depending on their electric charge and decay configurations.  Following similar considerations, colourless spin one resonances mixing with the SM electroweak (EW) gauge bosons, hereafter referred to as  $\rho$-mesons, are also requested around the TeV~scale. EW precision measurements from LEP experiments severely, yet indirectly, constrain additional gauge bosons  mixing with the $W$ and the $Z$. In a minimal CHM (MCHM) enjoying custodial symmetry~\cite{Agashe:2004rs,Agashe:2006at}, the most stringent limit arises mostly through the $S$ parameter~\cite{Peskin:1991sw}, which yields an indirect bound of $m_\rho\gtrsim 2\,$TeV (see \eg~Ref.~\cite{Contino:2015mha}). A scale comparable to that inferred from EW gauge boson contributions to the SM Higgs mass. Hence, naturalness, together with present collider constraints, typically point to a region of CHMs parameter space where $\rho$-mesons are likely to decay into on-shell top partners.

At the LHC, EW vector resonances are primarily sought  via narrow resonances searches.
Current constraints are driven by channels where the charged and neutral $\rho$-mesons decay into pairs of opposite charged leptons~\cite{Chatrchyan:2012oaa,Aad:2014cka}, 
and SM gauge bosons in the fully leptonic~\cite{Aad:2014pha}, fully hadronic~\cite{Khachatryan:2015bma} and semi-leptonic~\cite{Aad:2015ufa} final states. A reinterpretation of these searches in CHMs yields bounds which are typically around $2\,$TeV~\cite{Pappadopulo:2014qza,Thamm:2015zwa}. However, for these searches to be effective, spin one resonances have to directly decay with a substantial rate into SM states.  Yet, limits from such direct searches greatly weaken within a regime where $\rho$ resonances decay channels into top partner pairs kinematically open up. This is a direct consequence of significantly suppressed $\rho^{0,\pm}$ branching ratios (BRs)  into SM states  due to the strong intercomposite coupling constant $g_\rho\gg g_{\rm SM}$~\cite{Barducci:2012kk,Greco:2014aza}. In this case the strongest bound on EW spin one states is currently set by indirect limits from the $S$ parameter~\cite{Greco:2014aza}. Therefore, 
there is to date no direct probe of this kinematical regime favoured by naturalness.

 Nonetheless, we observe that $\rho$ decays into pairs of composite fermions give rise to peculiar final states which might as well compete with indirect constraints from LEP.
In particular, an interesting decay channel is that of the neutral $\rho$ into a pair of $X_{5/3}$ top partners with exotic electric charge $5/3$. This channel gives rise to a distinctive same sign dilepton (SS2$\ell$) signature which is currently exploited both by the ATLAS and CMS collaborations to bound the  $X_{5/3}$ mass through QCD-driven pair production~\cite{Chatrchyan:2013wfa,Aad:2015mba}.  We point out in this paper that  a simple reinterpretation of these analyses offers an interesting possibility to probe the otherwise elusive EW spin one resonances in a regime where they dominantly decay into top partners.
Similar approaches have been recently followed to constrain composite partners of the SM gluons~\cite{Azatov:2015xqa,Araque:2015cna}, as well as charged EW vector resonances~\cite{Vignaroli:2014bpa}.  We extend here the analyses of Refs.~\cite{Greco:2014aza} and~\cite{Vignaroli:2014bpa}, by recasting the SS2$\ell$ CMS analysis. In particular, through the inclusion of  all possible CHM contributions to the SS2$\ell$ final state, we show that the kinematical regime favoured by naturalness, as argued above, is already significantly constrained by the available $8\,$TeV LHC data, thus worsening the amount of fine-tuning associated with the top sector in MCHMs. We also argue about the usefulness of SS2$\ell$ searches to further constrain CHMs in future LHC upgrades.

 The remainder of the paper is organised as follows. We recall in Sec.~\ref{model} the effective Lagrangian capturing the main features of the simplest CHM, with particular emphasis on composite top partner interactions with EW spin one resonances. Our recast of the CMS analysis for the SS2$\ell$ signal is described in Sec.~\ref{recast}. The resulting reinterpretation of the 8$\,$TeV LHC data in a scenario where composite $\rho$-mesons dominantly decay into top partners is presented in Sec.~\ref{8tev}, while Sec.~\ref{13TeV} discusses the prospects at future 13$\,$TeV LHC runs. We conclude in Sec.~\ref{conc}.


\section{The Model}\label{model}

We consider the MCHM, with a strong sector globally invariant under SO(5)$\times$U(1)$_X$, where the   SO(5) factor spontaneously breaks down to SO(4) at a scale $f$. 
This is the most economical symmetry breaking pattern that provides 4 Goldstone bosons (GBs) and embeds a custodial symmetry in order to protect the EW $\rho$ parameter~\cite{Agashe:2003zs,Agashe:2004rs}.
Besides the GBs, we assume two sets of composite resonances below the cutoff $\Lambda\sim 4\pi f$ of the strong sector: one vector multiplet $\rho_\mu$, and one multiplet of vector-like fermions $\Psi$, which transform as $({\bf3},{\bf1})_0$ and $ ({\bf 2},{\bf 2})_{2/3}$, respectively, under the unbroken SO(4) (locally isomorphic to SU(2)$_L\times$SU(2)$_R$) and U(1)$_X$ symmetries.\footnote{For sake of concreteness we focus on an SU(2)$_L$ triplet of vector resonances.
Although their production cross sections are significantly smaller~\cite{Greco:2014aza}, our analysis also applies to vector resonances transforming non-trivially under SU(2)$_R\times$U(1)$_X$.}
We further assume that the right-handed top quark $t_R$ is a fully composite resonance of the strong sector\footnote{While this is not strictly speaking a necessary requirement, a fully composite $t_R$ is favored by a rather light $125\,$GeV Higgs boson~\cite{Panico:2012uw}, as well as allows for precise unification of the SM gauge couplings in the CH framework~\cite{Agashe:2005vg}.}, transforming as $({\bf 1},{\bf 1})_{2/3}$. The low energy Lagrangian below $\Lambda$ is determined by the SO(5)$/$SO(4) coset of symmetry breaking, and it has the form 
\be
\mathcal{L}_{\rm MCHM}=\mathcal{L}_{\rm el} +\mathcal{L}_{\rm co} +\mathcal{L}_{\rm mix}\,.
\label{eq:ltot}
\ee
The elementary sector Lagrangian is simply the SM Lagrangian without the Higgs and $t_R$ fields,\footnote{We leave QCD interactions implicit throughout the paper.} 
\be
\mathcal{L}_{\rm el} = i \bar q \slashed D q + i\psi\slashed D \psi-\frac{1}{4}(W_{\mu\nu}^a)^2-\frac{1}{4}B_{\mu\nu}^2\,,
\label{eq:lel}
\ee
where $q=(t_L,b_L)^T$ is the third generationi	 quarks doublet, $\psi$ collectively denotes the lighter SM fermions, $W_{\mu\nu}^a$ and $B_{\mu\nu}$ are the SU(2)$_L$ and U(1)$_Y$ field strength tensors respectively, and $D_\mu$ is the SM covariant derivative.
The standard Callan--Coleman--Wess--Zumino (CCWZ) formalism~\cite{Coleman:1969sm,Callan:1969sn} yields the following composite sector Lagrangian (see \eg~Ref.~\cite{Panico:2015jxa})
\bea
\mathcal{L}_{\rm co} &=& \frac{f^2}{4}d_\mu^2-\frac{1}{4} (\rho_{\mu\nu}^a)^2+\bar \Psi(i\slashed D +\slashed e-M_\Psi)\Psi +i\bar t_R\slashed D t_R\nonumber\\
&&+\frac{M_\rho^2}{2}\left(\rho_\mu^a-g_\rho^{-1}e_\mu^a\right)^2+\left( ic_1\Psi^i\slashed d^i t_R +\hc\right)\nonumber\\
&&+g_\rho c_3\bar \Psi\left(\slashed \rho^a-g_{\rho}^{-1}\slashed e^a\right)t^a\Psi\,,
\label{eq:lcom}
\eea
where $\rho_{\mu\nu}^a = \partial_\mu \rho^a_\nu -\partial_\nu \rho^a_\mu +g_\rho \epsilon_{abc} \rho^b_\mu\rho^c_\nu$, $t^a$ ($a=1,2,3$) are the SU(2)$_L$ generators, and
\beq\label{psi}
\Psi = \frac{1}{\sqrt{2}}\left(\begin{array}{c} iB_{-1/3}-iX_{5/3} \\ B_{-1/3}+X_{5/3} \\ iT_{2/3}+iX_{2/3}\\-T_{2/3}+X_{2/3}\end{array}\right)\,.
\eeq 
The subscript labeling the top partners in Eq.~\eqref{psi} indicates their electric charge.
$d_\mu$ and $e_\mu$ are CCWZ symbols ensuring the non-linear realisation of SO(5) in the effective theory. They are functions of the EW gauge fields, whose corresponding expressions are found in Appendix~\ref{CCWZ}. Finally the elementary/composite mixing terms are
\beq
\mathcal{L}_{\rm mix} = y_Lf\bar q_5^I U_{Ii} \Psi^i+y_L c_2f\bar q_5^I U_{I5}t_R+\hc\,,\label{Lmix}
\label{eq:lmix}
\label{eq:lmix}
\eeq
where $U_{IJ}$ ($I,J=1,\ldots,5)$ is a matrix built out of GBs fields (including the physical Higgs boson) and $q_5$ denote the SO(5) embedding of the elementary doublet~$q$,
 \beq
q_5=\frac{1}{\sqrt{2}}\left(ib_L, b_L,it_L,-t_L,0\right)^T\,. 
 \eeq
For simplicity we ignore possible mixing terms involving the light fermions $\psi$, which is justified in the partial compositeness paradigm~\cite{Kaplan:1991dc}, in particular if the strong sector is flavour anarchic~\cite{Grossman:1999ra,Huber:2000ie}.

 The model described above is a minimal realisation of the CHM paradigm with partial compositness. It contains in total 11 parameters, 4 of which must be fixed in order to reproduce $G_F$, $M_Z$, $\alpha$ and the top mass, leaving then 7 free parameters beyond the SM ones. Those are $f$, $g_\rho$, $c_{1,3}$, $y_L$, $M_\rho$ and $M_\Psi$.
In the following we will work under the assumption that $M_\rho=f g_\rho$, therefore directly linking the composite scale $f$ to the mass parameter of the EW resonances. 
Beside setting the interaction strength of the $\rho$-mesons with the composite fermions and SM third generation quarks, $g_\rho$ also controls the elementary/composite mixing that makes the composite vectors interact with SM leptons and light quarks, thus setting their production cross sections via Drell-Yan (DY) processes.
The $\rho$-couplings to third generation SM quarks and top partners is further regulated by $c_{1,3}$ and $y_L$, where the latter, which only affects the $T_{2/3}$ and $B_{-1/3}$ phenomenology, also sets the degree of compositness of the left-handed top quark.\\

The model characteristics relevant to collider phenomenology are as follows.
Before electroweak symmetry breaking (EWSB), the masses of the composite resonances read\footnote{ EWSB further mixes the composite and elementary states, thus inducing $\mathcal{O}(\xi\equiv v^2/f^2)$ corrections to the spectrum. While these corrections are parametrically small, $\xi\lesssim 0.1-0.3$ from LEP data~\cite{Giudice:2007fh}, they are not always negligible and therefore fully included in our numerical analysis. Furthermore, corrections of higher order in $g/g_\rho\ll1$ are understood in the expression of $m_\rho$.}
\beq
\begin{split}
& m_\rho \equiv m_{\rho^{0,\pm}}\simeq M_\rho\left(1+\frac{g^2}{8g_\rho^2}\right)\,,\\
& m_{X_{5/3},X_{2/3}} = M_\Psi\,, \\
& m_{T_{2/3},B_{-1/3}} = \sqrt{M_\Psi^2+ y_L^2 f^2}\,,
\end{split}
\eeq
where $g$ is the SU(2)$_L$ gauge coupling of the SM.
Besides modifications of the Higgs production and decay rates (see \eg~Ref.~\cite{Panico:2015jxa} and references therein for a recent review),  hallmark signatures of CHMs are the presence of light top partners below the TeV scale, and $\rho$ resonances with $m_\rho\gtap2\,$TeV. At the $8\,$TeV LHC the main production of top partners is via pair production mediated by QCD interactions, although at the present 13$\,$TeV centre-of-mass energy single top partner production also acts as an important probe~\cite{Atre:2011ae}. The Lagrangian of Eq.~\eqref{eq:lcom} and Eq.~\eqref{eq:lmix} yields the following approximate BRs pattern\footnote{ A simple way to obtain these BRs is through the equivalence theorem, whose use is justified by the significant mass splitting between top partners and SM states, $m_\Psi\gg m_{W,Z,h}$. These BRs could however significantly deviate from the above pattern in the presence of additional light fermionic resonances~\cite{Delaunay:2013pwa,Backovic:2014uma}.} for  top partners  
decays into  SM states (see \eg~Ref.~\cite{DeSimone:2012fs}) 
\beq 
\begin{split}
 & {\rm BR}(B_{-1/3}\rightarrow W^{-}t)\simeq {\rm BR}(X_{5/3} \rightarrow W^{+}t)= 100\%\\ 
 & {\rm BR}(X_{2/3}\rightarrow Zt)\simeq{\rm BR}(X_{2/3}\rightarrow Ht)\simeq50\% \\ 
 &  {\rm BR}(T_{2/3}\rightarrow Zt)\simeq{\rm BR}(T_{2/3}\rightarrow Ht)\simeq50\%\,.%
\end{split}
\label{eq:tprime-decay}
\eeq
Since the exotic $X_{5/3}$ quark can only decay via charged current interactions through the process $X_{5/3}\to(t\to~W^+ b)W^+$, leptonically decaying $W$ bosons give rise to a SS2$\ell$ signature. This final state configuration is subject to a significantly smaller SM backgrounds with respect to other top partner search channels, and it can thus be used as a powerful experimental probe. It is in fact currently exploited by both ATLAS and CMS collaborations to bound $X_{5/3}$ states pair produced through QCD interactions, setting a limit of $m_{X_{5/3}}\gtrsim 800\,$GeV at 95$\%$ confidence level (CL)~\cite{Chatrchyan:2013wfa,Aad:2015mba}. 

The  main production mode of EW spin one resonances at the LHC is via DY processes, while vector boson fusion production gives a negligible contribution to the total cross section.\footnote{For $g_\rho=2$, VBF accounts for $\mathcal{O}(1\%$) or less of the total cross section.} Experimental results of dilepton and diboson narrow resonances searches are usually expressed as limits on the $\rho$ production cross section times branching ratio into a given final state. These limits converts into bounds of $m_\rho\sim 1.5-2\,$TeV~\cite{Thamm:2015zwa}, under the assumption that the spin one states dominantly decay into SM final states. 
  However, in contrast with top partners, the BRs of the $\rho$ resonances strongly depend on the model parameters. In particular, once decays to top partners are kinematically allowed, {\it i.e.} when $m_{\rho}> 2 m_\Psi$, 
 direct decays into fermionic resonances pairs are significantly favoured relative to pure SM final states~\cite{Barducci:2012kk,Greco:2014aza}, essentially because of the large value of the intercomposite coupling constant $g_\rho\gg g$, see Eq.~\eqref{eq:lcom}. More quantitatively, the decay rates of $\rho^{0}\to X_{5/3}\bar X_{5/3},X_{2/3}\bar X_{2/3}$ and $\rho^+\to X_{5/3}\bar X_{2/3}$ quickly saturate  to $\sim$~60--70\%, thus rendering narrow resonance searches ineffective in bounding these states~\cite{Greco:2014aza}.
 We exploit this feature in the next section, where we derive the extend to which EW spin one resonances in this regime are already rather constrained by data collected during the LHC $8\,$TeV run.


\section{Recast of the CMS SS2$\ell$ analysis}\label{recast}

Our analysis is based on a recast version of the CMS search for $X_{5/3}$ in a SS2$\ell$ final state~\cite{Chatrchyan:2013wfa} implemented in the \ma5package~\cite{Conte:2012fm,Conte:2014zja,Dumont:2014tja} and publicly available on the Physics Analysis Database (PAD) web-page\footnote{\url{https://madanalysis.irmp.ucl.ac.be/wiki/PublicAnalysisDatabase}}. A full validation of the \ma5 implementation of this search is described in the provided validation note~\cite{MA5-CMS-B2G-12-012}. We give below a short summary of the results, referring to the full note for further details. For the signal prediction, we have used  an {\tt UFO} format~\cite{Degrande:2011ua} implementation of the model described in Sec.~\ref{model}, performed by the authors of Ref.~\cite{Greco:2014aza} through the {\tt Feynrules} package~\cite{Alloul:2013bka}, which has been made publicly available on the {\tt HEPMDB} website\footnote{\url{https://hepmdb.soton.ac.uk/hepmdb:1014.0179}}~\cite{Brooijmans:2012yi}. 

The CMS analysis selection requires isolated leptons, which are defined by computing the scalar sum of the $p_T$ of all neutral and charged reconstructed particles within a cone of size $\Delta R$ around the lepton momentum. This sum is then divided by the $p_T$ of the lepton, which is considered isolated if this ratio is below 0.15 (0.2) in a cone $\Delta R=0.3$ (0.4) for electrons (muons).
A category of loose leptons is also defined, where these ratios are increased to 0.60 (0.40) for electrons (muons).
Jets are reconstructed with \verb#FastJet#~\cite{Cacciari:2011ma}, via an anti-$k_T$~\cite{Cacciari:2008gp} algorithm, with a distance parameter of 0.5 and they are required to have $p_T>$ 30 GeV. In the experimental analysis, jet substructure algorithms to tag boosted jets from tops and $W$s decays, are also used. These features have not been implemented in the recast analysis. Nevertheless, as we will see, the $X_{5/3}$ mass bound obtained lies within a few \% from the official one.
\begin{figure}[tb]
\includegraphics[width=0.46\textwidth]{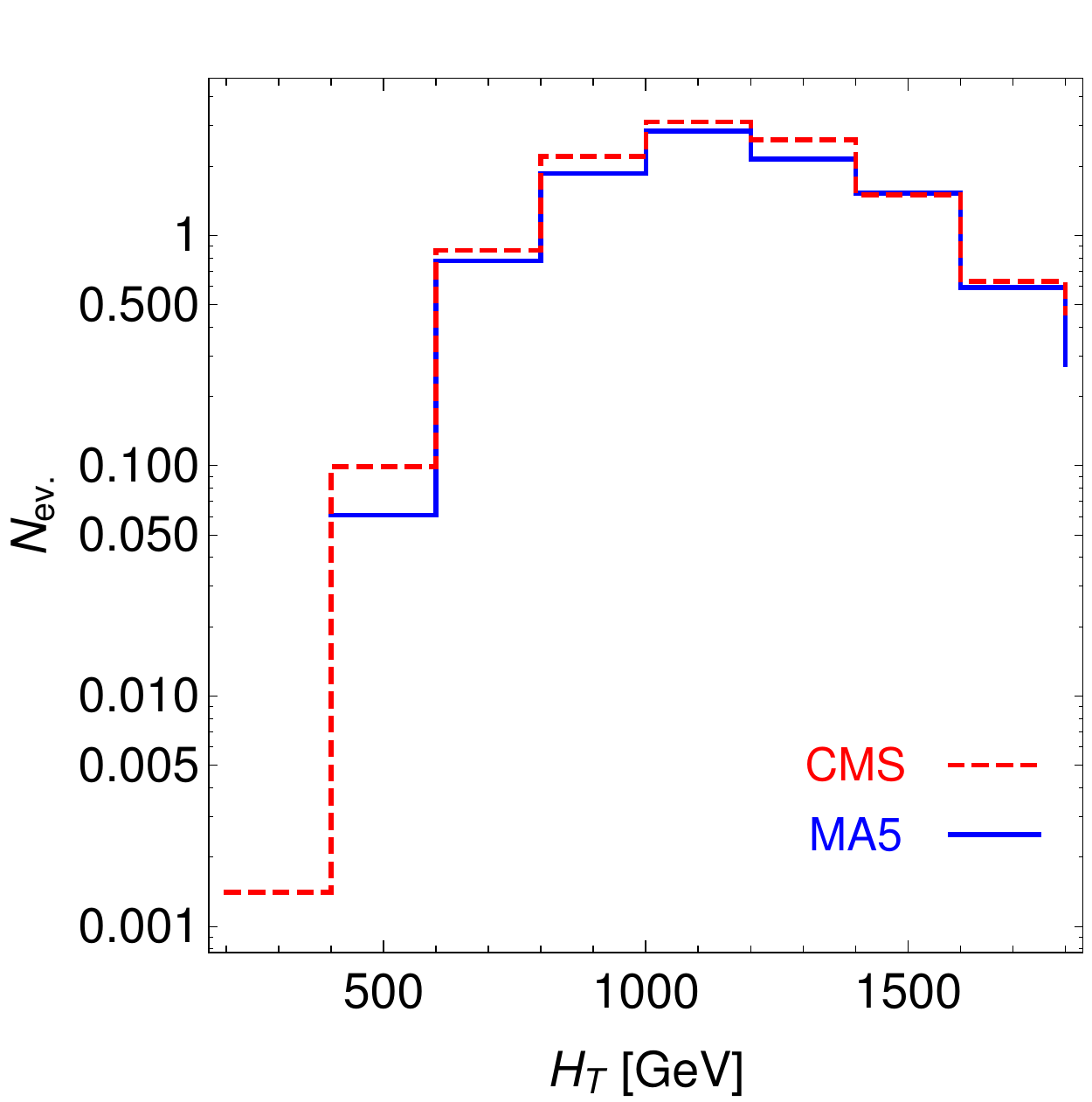}
\caption{$H_T$ distribution after the full event selection (without the $H_T$ requirement itself) for $m_{X_{5/3}}=800\,$GeV.}
\label{fig:valid1}
\end{figure}
\begin{figure}[tb]
\includegraphics[width=0.46\textwidth]{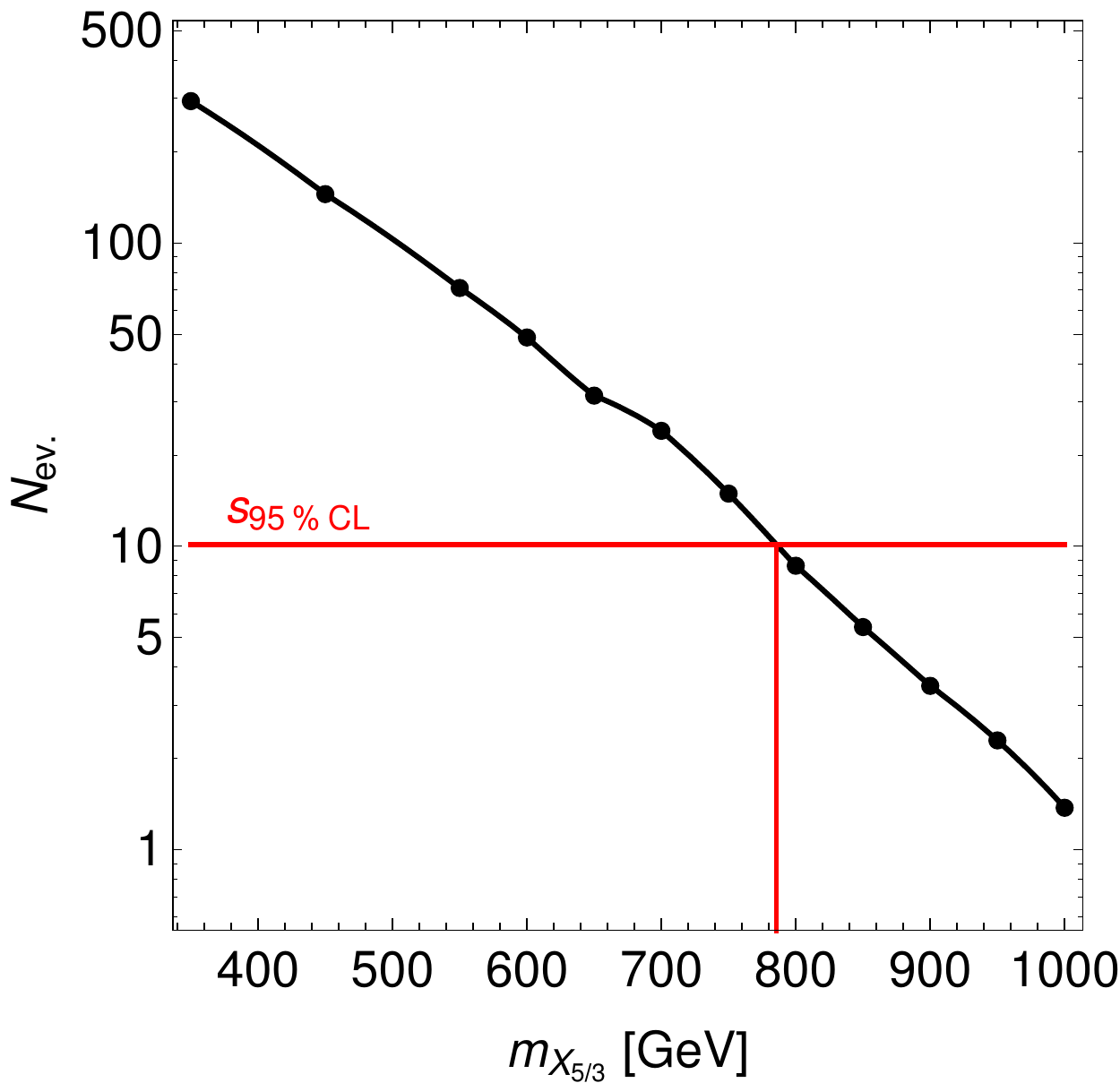}
\caption{Number of signal events surviving the selection cuts as a function of the $X_{5/3}$ mass. The  95\% CL excluded signal is shown in red. The dots correspond to the simulated mass points.}
\label{fig:valid2}
\end{figure}
The signal region is then defined by applying the following set of cuts:
\begin{itemize}
 \item at least two isolated same-sign leptons with $p_T>30\,$GeV,
 \item dilepton $Z$ boson veto: $|M_{ee}-M_Z|> 15\,$GeV, where $M_{ee}$ is the invariant mass of the same sign candidate electrons pair,
 \item trilepton $Z$ boson veto: $|M_{\ell\ell}-M_Z|> 15\,$ GeV, where $M_{\ell\ell}$ is the invariant mass of either one of the selected leptons and any other same flavour
opposite sign loose lepton with $p_T>15\,$GeV,
 \item $N_C\ge 7$, where $N_C$ is the number of constituents of the event,
 \item $H_T>900\,$GeV, where $H_T$ is the scalar sum of the $p_T$ of all the selected jets and leptons in the event.
\end{itemize}
In the validation note, several differential distributions for the signal have been checked and compared with the official CMS  results.
As an example, we report in Fig.~\ref{fig:valid1} the $H_T$ distribution after the application of all of the analysis cuts, with the exception of the $H_T$ requirement itself, for $m_{X_{5/3}}=$ 800 GeV. In the signal region  defined above, CMS observes  9 events, while the SM background hypothesis predicts 6.8$\pm$2.1 events. From these values, the CLs prescription~\cite{Read:2000ru,Read:2002hq} yields a 95\% CL exclusion limit for a signal rate giving 10.1 events.
Using our recast, this translate to a bound of $m_{X_{5/3}}\geq785\,$GeV, see Fig.~\ref{fig:valid2}, which is to be compared with the 800$\,$GeV bound from the original CMS analysis~\cite{Chatrchyan:2013wfa}.

\section{$m_\rho-m_{X_{5/3}}$ exclusion from $8\,$TeV data}
\label{8tev}

In our scenario, the main contributions to the SS2$\ell$ final state arise both from QCD pair production of the $X_{5/3}$ quark, and also from the $\rho^{0}$ and $\rho^+$ decays into a  $X_{5/3}\bar X_{5/3}$ and $ X_{5/3}\bar X_{2/3}$
final state, respectively. 
We have simulated with \verb#MG5_aMC v2.3.0#~\cite{Alwall:2014hca} DY production of $\rho^{0,\pm}$, with the subsequent decay of the spin one resonances in all possible final states combinations. This includes final state with a SM fermion and a top partner, which can contribute to the SS2$\ell$ final state when the decay into a top partner pair final state is kinematically forbidden. The same tool has also been used to simulate QCD pair production of a $X_{5/3}\bar X_{5/3}$ pair, up to two merged extra jets in the matrix element (ME). MLM matching scheme has been used~\cite{Mangano:2006rw,Alwall:2008qv}.
Parton showering, hadronisation and decay of unstable particles, including top partners, have been performed with \verb#PYTHIA v6.4#~\cite{Sjostrand:2006za}, while \verb#Delphes v3.2.0#~\cite{deFavereau:2013fsa} has been employed for a fast detector simulation. Jets have been reconstructed with \verb#FastJet#, via an anti-$k_T$ algorithm and a tuned CMS detector card suitable for performing a \ma5 analysis has been used.
Our signal samples have been generated for 
$M_\Psi\in[750,1000]\,$GeV and $M_{\rho}\in[1500,2500]\,$GeV, in steps of 50$\,$GeV and 250$\,$GeV, respectively, and fixing $g_\rho=2$, $f=M_\rho/g_\rho$ and $c_1=c_3=y_L=1$. We will however discuss in the following how our results are modified when deviating from these model parameters.
Finally, the samples have been passed through the \ma5 implementation of the CMS SS2$\ell$ analysis~\cite{MA5-CMS-B2G-12-012}.

\begin{center}
\begin{figure}[htb]
\includegraphics[width=0.46\textwidth]{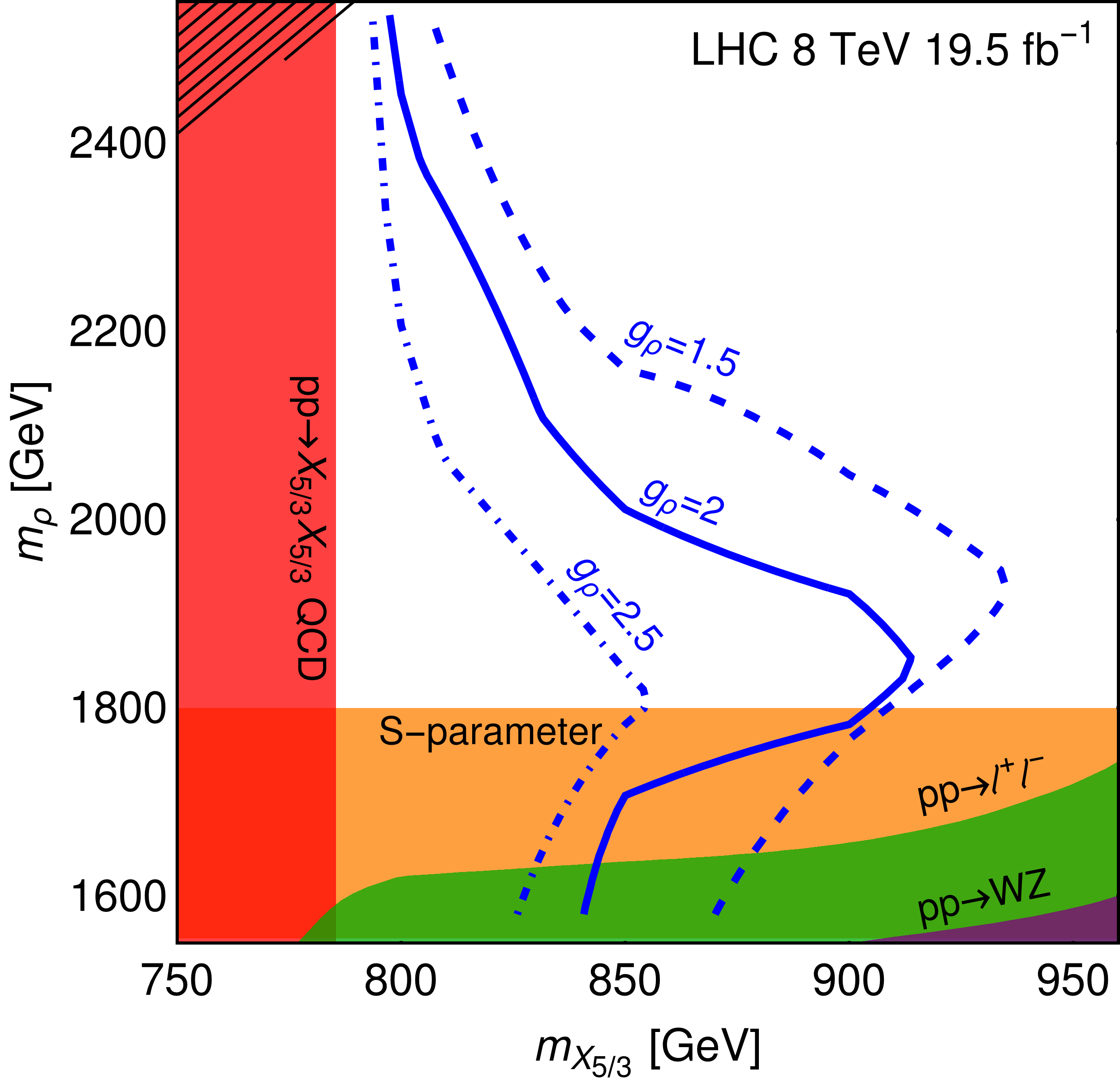}{}
\caption{95\% CL exclusion contours in the $m_{X_{5/3}}-m_\rho$ plane based on the recast described in Sec.~\ref{recast} at the 8 TeV LHC with 19.5 fb$^{-1}$ of integrated luminosity. Dashed, solid and dotted-dashed blue delineate the exclusion limits for $g_\rho=1.5,2,2.5$, respectively. The region to the left of each contour is excluded. 
The horizontal orange shaded band represents the indirect limit from the $S$ parameter, while the vertical red shaded band denotes the direct bound on $m_{X_{5/3}}$ from Ref.~\cite{Chatrchyan:2013wfa} assuming just QCD pair production. Green and purple shaded areas are the excluded regions from narrow resonance analysis, based on $\ell^+\ell^-$~\cite{Aad:2014cka} and $WZ$~\cite{Aad:2015ufa} final states, respectively, for $g_\rho=2$. The black hatched area in the top-left corner corresponds to the region where $\Gamma_\rho/m_\rho> 20\%$. See text for details.  }
\label{fig:8TeV}
\end{figure}  
\end{center}

Our results are shown in Fig.~\ref{fig:8TeV}, where the blue solid line delineates the excluded region at 95\% CL in the $m_\rho-m_{X_{5/3}}$ plane for $g_\rho=2$. The obtained limit clearly shows that the presence of $\rho$ resonances, both charged and neutral, decaying into top partners, improves upon the pure QCD exclusion limit set by the CMS analysis on the $X_{5/3}$ quark mass (denoted by the vertical red-shaded band in Fig.~\ref{fig:8TeV}). This limit is recovered with no ambiguity when $m_\rho$ increases. Most notably, the bound obtained on $m_\rho$ is stronger than the indirect one from the $S$ parameter~\cite{Greco:2014aza} (represented by the horizontal orange-shaded band). 
For the sake of comparison we also show  in Fig.~\ref{fig:8TeV} the ATLAS reach of the narrow resonance searches in $\ell^+\ell^-$~\cite{Aad:2014cka} and $WZ\to \ell E_T^{\rm miss} j j$~\cite{Aad:2015ufa} final states (green and purple shaded regions, respectively), also assuming  $g_\rho=2$. As argued above, these analyses quickly loose sensitivity above the threshold $m_\rho=2 m_{X_{5/3}}$, making the $\rho$ resonances escape the current LHC limits.
Finally, the black-hatched area represents the region where $\Gamma_\rho/m_\rho\geq~20\%$.
Note that the bulk of the exclusion derived from our recast of the SS2$\ell$ search lies in a region of relatively moderate resonance width, therefore justifying  the use of a Breigth Wigner (BW) propagator to simulate the signal. For higher values of this ratio, a full momentum-dependent width  ought to be used in the resonance propagator, while no reliable prediction can be calculated for $\Gamma_\rho/m_\rho\gtrsim 1$ as one enters a strongly-coupled regime.

The above results have been derived assuming $g_\rho=2$, $f=M_\rho/g_\rho$ and $c_1=c_3=y_L=1$. However, a slight modification of these parameters moderately alters the exclusion reach of the SS2$\ell$ search. While the $\rho$ production cross section is  only function of the composite coupling $g_\rho$, its decay rates are also function of $c_1$, $c_3$ and $y_L$.
The parameter $c_1$ has a negligible impact on the pure composite decays, while $c_3$, $g_\rho$ and, to a lesser extend, $y_L$ can have a stronger influence.
Nevertheless, above the threshold $m_\rho>2 M_\Psi$, the $\rho^0$ BR into the $X_{5/3}\bar X_{5/3}+X_{2/3}\bar X_{2/3}$ final state rapidly saturates and a modification of $g_\rho$ and/or $c_3$  mainly changes the width of the resonances. For the sake of completeness, we show in Fig.~\ref{fig:8TeV} the 95\% CL exclusion limits for $g_\rho=1.5$ (blue dashed) and $g_\rho=2.5$ (blue dot-dashed), with all the other parameters unchanged, with the exception of $c_3$, which is rescaled as $2/g_\rho$ in order to keep the $\Gamma_{\rho}/m_{\rho}$ ratio  constant for fixed $m_\rho$ and $M_\Psi$ values. A reduction of $g_\rho$ results in a stronger exclusion, due to the enhanced elementary/composite mixing with the SM gauge bosons which leads to a higher production cross section. Conversely, the exclusion reach of the SS$2\ell$ search is strongly reduced if $g_\rho$ is slightly increased from our initial choice $g_\rho=2$.\footnote{ Changing $g_\rho$ also alters the exclusion reach of the narrow resonance searches, which improves for smaller values of $g_\rho$. We however verified that these analyses always remain significantly less effective than our recast of the SS$2\ell$ search in bounding the region of parameter space where $m_\rho>2m_{X_{5/3}}$.} In the presence of the $\rho$ resonance the reach on the $X_{5/3}$ mass increases up to $\sim930\,$GeV for $g_\rho=1.5$ and $m_\rho=1.9\,$TeV, which is stronger than that of pure QCD production by $\sim20\%$. This results in a more significant fine tuning associated with the top sector. $g_\rho=1.5$ corresponds to a value of $\xi\simeq0.04$. For this small value of $\xi$, top partner masses favoured by the 125$\,$GeV Higgs can be higher~\cite{Matsedonskyi:2012ym} and above the threshold for $\rho$-mediated pair production.
Nevertheless, naturalness considerations still require light top partners below $\sim1\,$TeV, a regime where the above analysis has a high sensitivity.

\section{$13\,$TeV LHC projections}\label{13TeV}

We now evaluate the sensitivity to the same scenario of an upgraded SS2$\ell$ search at the 13$\,$TeV LHC runs. We begin with deriving the projected limits on the $X_{5/3}$ mass assuming only QCD pair production. We used the selection criteria proposed in Ref.~\cite{Avetisyan:2013rca} and, as for the case of the 8 TeV analysis, we did not attempt to use any jet substructure technique. With respect to the 8 TeV case, the following cuts have been modified or added:
\begin{itemize}
\item at least two isolated same sign leptons with $p_T>80\,$GeV,
\item leading and second leading jets with $p_T>150\,$GeV and 50$\,$GeV respectively,
\item $E_T^{\rm miss}> 100\,$GeV,
\item $H_T>1500\,$GeV,
\item $S_T=H_T+E_T^{\rm miss}>2000\,$GeV.
\end{itemize}
Signal samples for QCD pair production of extra quarks (again up to 2 merged extra jets in the ME), with $m_{X_{5/3}}\in[1000,2000]\,$GeV have been generated in steps of 100$\,$GeV, together with the main SS2$\ell$ SM backgrounds, namely $t\bar t W$, $t\bar t Z$, $WW$, $WZ$ and $WWW$.
The normalisation of the $X_{5/3}$ QCD pair production cross section has been computed with \verb#Hathor v.2.0#~\cite{Aliev:2010zk}, while for the SM backgrounds the leading order predictions were computed with \verb#MG5_aMC v2.3.0#.
For the background processes we report the cross sections and acceptance times efficiencies values in Tab.~\ref{tab:13tevbkg}, from which it is straightforward to calculate the number of signal events excluded at 95\%~CL, using the CLs prescription.

\begin{table}[h!]
\begin{center}
\begin{tabular}{ l | c | c } 
\hline
Process         		&~~~ $\sigma$ [fb]~~~	   	&~~ $A\cdot\epsilon\  [\times  10^{-5}]$~~ \\
\hline
\hline
$pp\to t\bar t W (0j+1 j)$      & 483.0        			& 6.29   \\
$pp\to t\bar t Z (0j+1 j)$      & 633.0        			& 1.19     \\ 
$pp\to W^+ W^+ jj  $      	& 187.3        			& 2.60   \\
$pp\to WZj (W,Z\to \ell \nu,\ell^+\ell^-)  $   & 59.0        			& 3.90   \\
$pp\to WWWj (0j+1j)  $      	& 166.2        			& 1.35   \\ 
\hline
\end{tabular}
\end{center}
\caption{Cross section and acceptance times efficiency values for the main SM backgrounds contributing to the SS2$\ell$ final state. The $WZj$ sample has been generated with a generator level cut on the leading jet $p_T$ of 120 GeV.}
\label{tab:13tevbkg}
\end{table}
\begin{figure}[tb]
\includegraphics[width=0.46\textwidth]{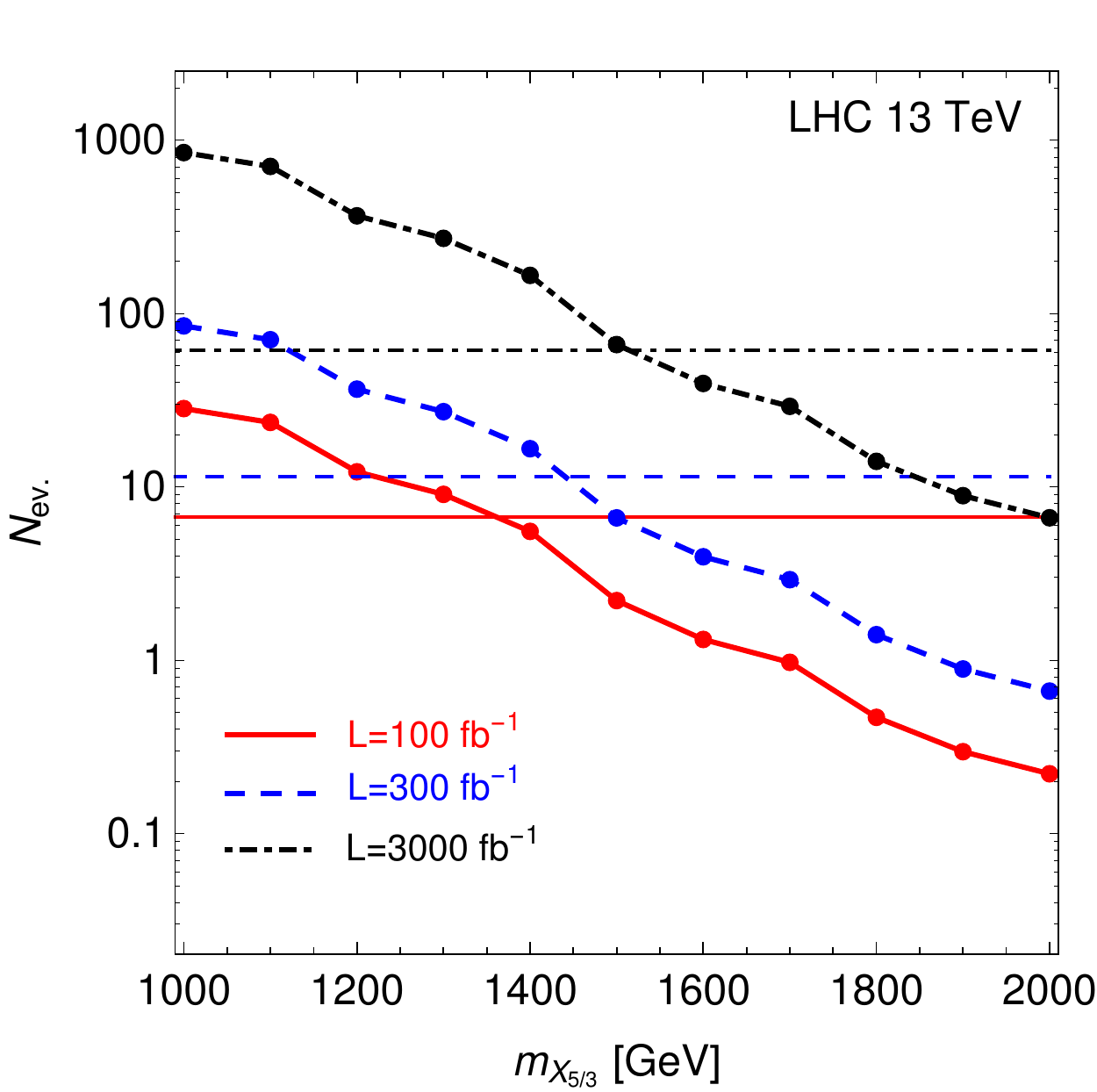}
\caption{Number of signal events surviving the selection cuts as a function of the $X_{5/3}$ mass for the 13 TeV LHC with 100, 300 and 3000$\,$fb$^{-1}$ (red, blue and black curves, respectively). The dots correspond to the simulated mass points. Also shown are the 95\% CL excluded signal rates (horizontal lines).}
\label{fig:13qcd}
\end{figure} 
\begin{table}[h!]
\begin{tabular}{ l || c | c | c } 
\hline
			&~~ $N_{\rm bkg.}$ ~~	 &~~~ $s_{95\% \rm CL}$~~~ & $m_{X_{5/3}}$[GeV]  \\
\hline
\hline
$\mathcal{L}=100$ fb$^{-1}$      & 4.73     			 & 6.7  			& 1366 \\
$\mathcal{L}=300$ fb$^{-1}$     & 14.20      			 & 11.4 			& 1452 \\
$\mathcal{L}=3000$ fb$^{-1}$     & 142.98      		 & 61.2 			& 1518 \\
\hline
\end{tabular}
\caption{Background yields, 95\% CL excluded signal rates and projections on the $X_{5/3}$ mass reach, for three benchmark values of integrated luminosity for the 13 TeV LHC: 100, 300 and 3000 fb$^{-1}$.}
\label{tab:13qcd}
\end{table}
We report these values in Tab.~\ref{tab:13qcd} for the three projected milestones of integrated luminosity achievable at the 13 TeV LHC, namely $\mathcal{L}=100,300,3000\,$fb$^{-1}$. (We  assumed a 20\% uncertainty on the background determination.) Also reported in Tab.~\ref{tab:13qcd} are  the projected exclusions limits at 95\% CL on the $X_{5/3}$ mass, which ranges from 1360 to 1520$\,$GeV, depending on the LHC luminosity. Figure~\ref{fig:13qcd} further shows the SS2$\ell$ final event rates as a function of the $X_{5/3}$ mass, as well as the excluded signal rates. These results are in good  agreement with previous studies, see \eg~Ref.~\cite{Avetisyan:2013rca}.\\

Using the same procedure adopted for reinterpreting the 8$\,$TeV data, we now illustrate the reach of the 13$\,$TeV run of the LHC on the full CHM parameter space, {\it i.e.} in the presence of light EW spin one resonances.
Signal samples corresponding to $M_{\rho}\in[2500,4000]\,$GeV and $M_\Psi\in[1000,2000]\,$GeV have been 
generated in steps of 250 and 100$\,$GeV respectively, while fixing the other model parameters to $g_\rho=2$, $f=M_\rho/g_\rho$ and $c_1=c_3=y_L=1$. Similarly to the 8 TeV analysis, $\mathcal{O}(1)$ modifications of the model parameters will lead to moderate distortions of the exclusion limits.
\begin{figure}[tb]
\includegraphics[width=0.46\textwidth]{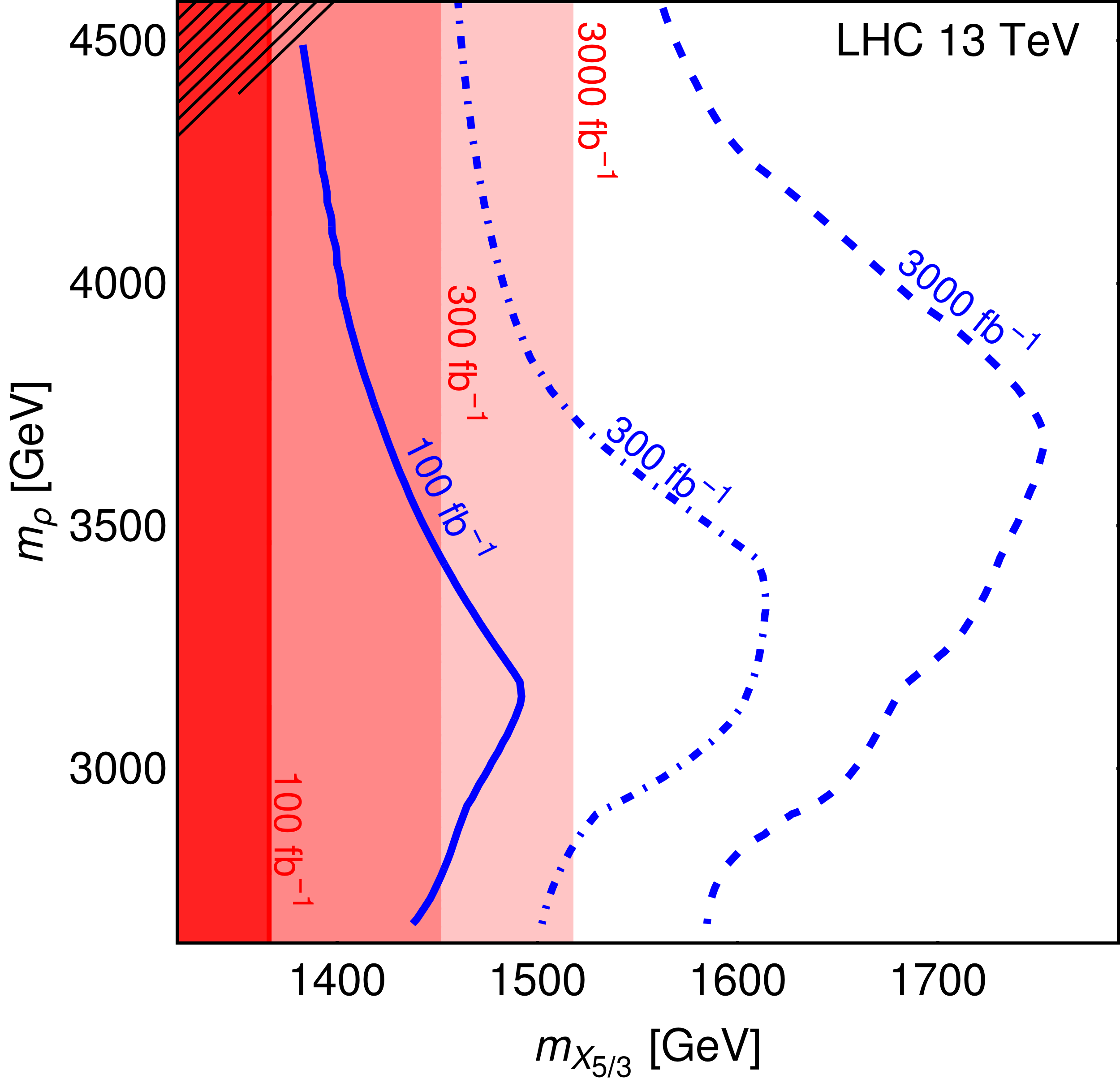}
\caption{95\% CL projected exclusion limits in the $m_{X_{5/3}}-m_\rho$ plane at the 13 TeV LHC for $g_\rho=2$. Integrated luminosities of 100$\,$fb$^{-1}$ (solid), 300$\,$fb$^{-1}$ (dotted-dashed) and 3000$\,$fb$^{-1}$ (dashed) are assumed.
The red shaded areas represent the limits assuming just QCD pair production of the $X_{5/3}$ quark, while the black hatched area in the top-left corner corresponds to the region $\Gamma_\rho/m_\rho>$ 20\%.}
\label{fig:lhc13_100-3000}
\end{figure} 
We then show  in Fig.~\ref{fig:lhc13_100-3000} the excluded regions of the $m_\rho-m_{X_{5/3}}$ plane for $\mathcal L=100,300,3000\,$fb$^{-1}$. It appears clearly that already with 100 fb$^{-1}$ at 13$\,$TeV, the LHC will be able to greatly improve upon the 8$\,$TeV exclusion reach, up to $m_\rho\simeq 3.2\,$TeV for  $m_{X_{5/3}}\simeq1.5\,$TeV. Note that the corresponding exclusion reach from QCD pair production only is $m_{X_{5/3}}\simeq1.4\,$TeV, while it reaches $m_{X_{5/3}}\simeq1.5\,$TeV with 3$\,$ab$^{-1}$. At the end of the LHC program ($\mathcal{L}=3\,$ab$^{-1}$) the SS2$\ell$ coverage will extend up to $m_{\rho}\simeq3.7\,$TeV with $m_{X_{5/3}}\simeq 1.7\,$TeV. Again, note that the bulk of the exclusion lies in a region where $\Gamma_{\rho}/m_{\rho}$ remains below 20\%, thus maintaining the good reliability of our analysis.
Finally we would like to comment about the implication of $\rho$ mediated top partners production modes in the case of an observed excess in the SS2$\ell$ channel at the 13$\,$TeV LHC.
While kinematic distributions for final state objects from $\rho$ and QCD production can in principle differ, the two production modes were found to be almost indistinguishable after detector resolution effects are taken into account~\cite{Araque:2015cna}.
 Therefore, only a full reconstruction of the top partner pair invariant mass could unveil the presence of $\rho$-mediated production modes.


\section{Conclusions}\label{conc}

In this paper we analysed the CHM paradigm in a regime where EW vector resonances are present in the effective theory and significantly decay into pairs of top partners, a region of parameter space well motivated by naturalness consideration and EW precision tests. While LHC searches for narrow resonances are unable to bound the composite vector states in this regime, 
we showed that experimental analyses originally designed to search for QCD pair produced top partners, namely  SS2$\ell$ searches, have a significant coverage of this region of parameter space already with the 8 TeV LHC data. We stress that TeV-scale EW resonances must be present below the cut-off of the strong dynamics in order to balance the SM gauge boson source of EW fine-tuning. These resonances then enhance the production cross section for top partners. Through a recast of the existing CMS SS2$\ell$ search, we quantified the extent to which the experimental
reach for $X_{5/3}$ top partners improves. This results in a
appreciable worsening of the CHMs fine-tuning associated
with the top sector. Finally, we discussed the prospect to further improve these results with the 13 TeV stage of the LHC.

\section*{Acknowledgments}
The authors thank the CMS collaboration, in particular Aram Avetisyan and Devdatta Majumder, for providing details about the CMS SS2$\ell$ analysis.
They also thank Genevi\`eve B\'elanger for useful discussion during the completion of this work. The work of CD is supported by the ``Investissements d'avenir, Labex ENIGMASS''.

\appendix

\section{SO(5)$/$SO(4) notations}\label{CCWZ}
We define the 10 generators of $SO(5)$ in the fundamental representation as
\beq
(T^\alpha_{L})_{IJ} = -\frac{i}{2}\left[\frac{1}{2}\varepsilon^{\alpha\beta\gamma}
\left(\delta_I^\beta \delta_J^\gamma - \delta_J^\beta \delta_I^\gamma\right) +
\left(\delta_I^\alpha \delta_J^4 - \delta_J^\alpha \delta_I^4\right)\right]\,,\nonumber
\eeq
\beq
(T^\alpha_{R})_{IJ} = -\frac{i}{2}\left[\frac{1}{2}\varepsilon^{\alpha\beta\gamma}
\left(\delta_I^\beta \delta_J^\gamma - \delta_J^\beta \delta_I^\gamma\right) -
\left(\delta_I^\alpha \delta_J^4 - \delta_J^\alpha \delta_I^4\right)\right]\,,\nonumber
\label{eq:SO4_gen}
\eeq
\begin{equation}
T^{i}_{IJ} = -\frac{i}{\sqrt{2}}\left(\delta_I^{i} \delta_J^5 - \delta_J^{i} \delta_I^5\right)\,,
\label{eq:SO5/SO4_gen}
\end{equation}
where $I,J=1,\ldots ,5$. $T_{L,R}^\alpha$ ($\alpha=1,2,3$) are the unbroken generators of SO(4)$\simeq$SU(2)$_L\times$SU(2)$_R$, while $T^i$ ($i=1,\ldots,4$) are the broken generators associated with the coset SO(5)$/$SO(4); They are all normalised such that $\tr(T^AT^B)=\delta^{AB}$.
It is convenient to collectively define $T^{\bar a}$ ($\bar a=1,\ldots,6)$, where $T^{1,2,3} = T_L^{1,2,3}$ and $T^{4,5,6}=T_R^{1,2,3}$, which, in the basis of Eq.~\eqref{eq:SO4_gen}, read
\beq
T^{\bar a}=\left(\begin{array}{cc}t^{\bar a} &0 \\ 0 &0 \end{array}\right)\,,
\eeq 
where $t^{\bar a}$ are the 6 SO(4) generators in the fundamental representation. We also use the notation $t^a$ for $t^{1,2,3}$ to explicitly refer to the generators of the SU(2)$_L$ subgroup of SO(4), as well $T^a$ for their embedding in SO(5).
 
The matrix $U$ of Eq.~\eqref{Lmix} depends on the GB fields $\Pi_i$ as 
\beq
U &=& \exp\left[i \frac{\sqrt{2}}{f} \Pi_{i} T^{i}\right]\nonumber\\
&=&\left(\begin{matrix}
\displaystyle \textbf{1}_{4\times 4}-{\vec\Pi\vec\Pi^T\over \Pi^2} \left(1-\cos{\Pi\over f}\right)\hspace{1em}&
\displaystyle {\vec\Pi\over \Pi}\sin{\Pi\over f}\\
\rule{0pt}{2.em}\displaystyle -{\vec\Pi^T\over \Pi}\sin{\Pi\over f}&\displaystyle \cos{\Pi\over f}\\
\end{matrix}\right)\,,\label{gmatr}
\eeq
where $\vec{\Pi}\equiv (\Pi_1,\Pi_2,\Pi_3,\Pi_4)^T$ and $\Pi\equiv\sqrt{\vec{\Pi}\cdot\vec{\Pi}}$.
In unitary gauge, with $\Pi_{1,2,3}=0$ and $\Pi_4=\bar h\equiv v+h$, Eq.~\eqref{gmatr}
 becomes 
\beq
U=\left(\begin{array}{ccc}
                \mathbf{1}_{3\times 3} &  & \\
		&  \cos{\bar h/f} & \sin{\bar h/f} \\
		  & -\sin{\bar h/f} & \cos{\bar h/f}
               \end{array}\right)\,.
\label{gmatrU}
\eeq

The $d_\mu$ and $e_\mu$ CCWZ symbols are defined by $d_\mu=d_\mu^iT^i$ and $e_\mu=e_\mu^{\bar a} t^{\bar a}$ with
\beq
d_\mu^i &=&\sqrt{2}\left(\frac{1}{f}-\frac{\sin{\frac{\Pi}{f}}}\Pi\right)\frac{\left(\vec{\Pi}\cdot \nabla_\mu\vec{\Pi}\right)}{\Pi^2}\Pi^{i}\nonumber\\
&&+\sqrt{2}\,\frac{\sin{\frac{\Pi}{f}}}\Pi\nabla_\mu\Pi^{i}\,,\\
e_\mu^{\bar a}&=&-A_\mu^{\bar a}+4i\sin^2\left({\frac{\Pi}{2f}}\right)\ \frac{\vec{\Pi}^T t^{\bar a}\nabla_\mu\vec{\Pi}}{\Pi^2}\,.
\eeq
$\nabla_\mu \Pi$ is the covariant derivative of the Goldstone fields
\beq
\nabla_\mu\Pi^{i}=\partial_\mu\Pi^{i}-i A_\mu^{\bar a}\left(t^{\bar a}\right)^{i}_{\  j}\Pi^{ j}\,,
\eeq 
where $A^{\bar a}_\mu$ contains the  elementary SM EW gauge fields written, embedded in SO(5) as
\beq
A_\mu^{\bar a} T^{\bar a} =
gW^a_\mu T^a + g' B_\mu T_R^3\,,
\label{gfd}
\eeq
with $W^1_\mu=(W_\mu^++W^-_\mu)/\sqrt{2}$, $W^2_\mu=i(W_\mu^+-W_\mu^-)/\sqrt{2}$, $W_\mu^3= c_W Z_\mu+s_WA_\mu$ and $B_\mu=-s_W Z_\mu+c_W A_\mu$, $c_W$ $(s_W)$ being the cosine (sine) of the weak mixing angle. Note that $d_\mu$ and $e_\mu$ transforms under SO(4) as a fundamental and an adjoint representation, respectively.
Their components in unitary gauge read
\beq
d_\mu^{1,2}&=&-\frac{g W_\mu^{1,2}}{\sqrt{2}}\sin\frac{\bar h}{f}\,,\nonumber\\
d_\mu^3&=&\frac{g'B_\mu-gW_\mu^3}{\sqrt{2}}\sin\frac{\bar h}{f}\,,\nonumber\\
d_\mu^4&=&\frac{\sqrt{2}}{f}\partial_\mu h\,,
\eeq
and
\beq
e_\mu^{1,2}&=&-gW_\mu^{1,2}\cos^2\frac{\bar h}{2f}\,,\nonumber\\
e_\mu^3&=&- gW^{3}_\mu\cos^2\frac{\bar{h}}{2f}- g'B_\mu\sin^2\frac{\bar{h}}{2f}\,,\nonumber\\
e_\mu^{4,5}&=&-gW_\mu^{1,2}\sin^2\frac{\bar h}{2f}\,,\nonumber\\
e_\mu^6&=&-g' B_\mu\cos^2\frac{\bar{h}}{2f} -gW^{3}_\mu\sin^2\frac{\bar{h}}{2f}\,.
\eeq 

\section{The role of the SM mediated processes}
\label{sec:interf}

Processes mediated by SM EW gauge bosons, as well as their interference with the $\rho$-mediated processes, also contribute to the SS2$\ell$ final state.  We only aim here at estimating the SM contribution to the SS2$\ell$ signal for a set of representative benchmark points. We expect the impact of the SM processes to be small and to strongly depend on the resonance width, which controls the overlap between the $\rho$ and $W,Z$ contributions. We therefore chose to focus on the following points in the $m_{\rho}-m_{X_{5/3}}$ plane, namely ($m_{X_{5/3}}$,$m_{\rho}$)=
(1.7,3.25), (1.6,3.25), (1.5,3.5) and $(1.5,4.25)\,$TeV, while keeping $g_\rho=2$, $f=M_\rho/g_\rho$ and $c_1=c_3=y_L=1$. These points are roughly aligned along the expected 95$\%$ CL exclusion line of Fig.~\ref{fig:lhc13_100-3000}, corresponding to 300$\,$fb$^{-1}$ of integrated luminosity at 13 TeV. For the above points $\Gamma_{\rho}/m_{\rho}$ ranges from 6 to 20\%. We then simulated for each point signal samples assuming the full ME, \ie~including SM EW gauge bosons in the $s$-channel, and compared the total number of events passing the 13 TeV selection with the ones obtained from the $\rho$ resonance contribution only, assuming in both cases an integrated luminosity of 300 fb$^{-1}$. These numbers, along with the pure QCD contribution to the signal, are reported in Tab.~\ref{tab:interf}.
Although we focused only on a few benchmark points, these numbers still provide useful information.
 First of all, note that the inclusion of the SM processes can both increase and decrease the number of signal events passing the selection cuts, depending on the relative importance of the interference term. Then, the SM effect is more pronounced the wider the resonance, due to the larger overlap between the SM and $\rho$-mediated amplitudes,  enhancing the pure composite resonance contribution up to $\sim50\%$ for the last benchmark point with $\Gamma_\rho/m_\rho=20\%$. Yet, the exclusion reach remains approximately the same (albeit slightly higher) since QCD processes still clearly dominate the signal cross section in this region. In the opposite regime where the $\rho$ is relatively narrow, as in the first two benchmark points, EW production however tends to dominate over QCD. However, in this case, the smaller width suppresses the interference term. In turns, the SM contributions only marginally modify the total cross section, thus leaving the exclusion limit practically unchanged. We conclude that the results presented in the main text, which do not include the SM contributions, are already accurate enough given the other sources of uncertainty in our analysis.
\begin{table}[!h]
\begin{tabular}{ c | c | c || c | c | c | c | c} 
\hline
$m_{X_{5/3}}$&$m_{\rho}$  &$\Gamma_\rho/m_\rho$	& QCD	        & EW-$\rho$ 	& EW-full & CL-$\rho$ & CL-full\\
\hline
\hline
1.7 & 3.25    		  & 6\%	& 2.9		&  5.0			& 4.7 & $1.5\sigma$ & $1.4\sigma$\\
1.6 &3.25     	 &  11\%	 & 3.9	   	&  6.6			& 5.9 & $2.0\sigma$ &$1.8\sigma$\\
1.5&3.5  		  & 15\%	& 6.6		&  5.0			& 6.8 & $2.1\sigma$&	$2.4\sigma$\\
1.5&4.25  		  & 20\%	& 6.6		&  1.7			& 2.5	& $1.6\sigma$ &$1.7\sigma$\\
\hline
\end{tabular}
\caption{Number of events after the 13 TeV selection for QCD pair production of $X_{5/3}X_{5/3}$, EW  production from $\rho$ resonances only (EW-$\rho$) and full EW production including pure SM and interference contributions (EW-full). Integrated luminosity is 300 fb$^{-1}$. Masses are in units of TeV. CL-$\rho$ and CL-full denote the confidence level (in units of the standard deviation $\sigma$) with which each benchmark point is excluded assuming QCD+EW-$\rho$ and QCD+EW-full, respectively.}
\label{tab:interf}
\end{table}

\bibliographystyle{apsrev}
\bibliography{widerho}

\end{document}